\documentclass[aip,apl,reprint,twocolumn]{revtex4-1}
\usepackage{graphicx, amsmath, amssymb}

\usepackage{dcolumn}
\usepackage{bm}

\begin{document}

\title{Sub-THz thermally activated-electrical conductivity of CdS thin films.} 

\author{Rezwanur Rahman and John A. Scales}
\affiliation{Department of Physics, Colorado School of Mines, Golden, CO 80401-1887, USA}

\begin{abstract}

    The electrical conductivity of a CdS thin film, controlled by grain structures is
    essential to enhance its photoconductivity to be able to be fit as a window material in
    CdS/CdTe heterojunction solar cell. In order to characterize a thin film, 
    electromagnetically, we employed an open cavity resonator with a sub-millimeter VNA (Vector Network Analyzer). Our technique 
    is capable of measuring complex dielectric permittivity, $\tilde{\epsilon}$, of a photovoltaic film
    as thin as 0.1 $\mu$m. We measured the real part of the complex dielectric permittivity, $\epsilon_{re}$, 
    and electrical conductivity, $\sigma_{re}$ (derived from the imaginary part, $\epsilon_{im}$), 
    of unannealed and annealed CdS films with thicknesses $\sim$ 0.15 $\mu$m
    on $\sim$ 3 mm thick-borosillicate glass substrates, at room temperature. We obtain the (thermally activated) electrical conductivity
    between 100 and 312 GHz, which is less in annealed samples than in unannealed one by $\sim$ 2 orders of magnitude.
    Contrary to our expectations, the carrier concentrations extracted from these data by fitting a Drude model, are $\sim$ 10$^{16}$ cm$^{-3}$ (unannealed)
    and $\sim$ 10$^{14}$ cm$^{-3}$ (annealed). We investigate the connection between grain size and carrier concentration.

\end{abstract}

\maketitle

  Cadmium sulphide (CdS) is an integral component of CdTe/CdS heterostructured-thin solar cell where it is used as a window material
for its better photoconductivity, optolelectronic properties, and flexibilty.\citep{TLC91, TA97, MTS00} 
Its photoconductivity sensitivity depends on its 
electronic properties. CdS is usually layered
with cadmium telluride (CdTe) to make efficient solar cells.\citep{HRM06,SW01}
Cadmium sulphide (CdS) contains many intrinsic defect-states which affect its electronic conductivity profile. 
The grain structure can also plays a role. The microstructures of polycrystalline materials are composed of 
different oriented-grains.\citep{RB92,RB60} The grain boundaries are characterized by a potential barriers.
The photoexcitation can impact transport through these grain boundaries by various mechanisms. The electrical conductivity
is the key property in all these phenomena. This photovoltaic material can be grown with many different techniques, and 
dielectric properties are growth-technique dependent.

   The annealing-effects (in vacuum and under argon atmosphere) on the electrical and optical conductivity of CdS (deposited
by chemical bath deposition method) have been studied in the temperature range 200-400 K.\citep{HO01} 
In the article by \citet{HO01}, for temperatures $>$ 300 K, the conduction is explained by the traps at the grain boundaries for grains partially depleted. On the other hand,
for temperatures $<$ 300 K, the conduction results from a hopping due to localized states.

   In order to evaluate the quality of materials used in CdS/CdTe photovoltaic heterojunctions, the complex dielectric properties 
of polycrystalline CdS, formed into a thin film by sputter-deposition onto Silicon (Si) wafers, were investigated by \textit{in situ} 
spectroscopic ellipsometry.\citep{JL10} The authors focused on understanding the excited carriers by fitting their data to carrier scattering model. 
This is in optical frequency range (0.75 to 6.5 eV).

   \citet{LL72I} illustrated the relationship between the electrical properties, i.e. conductivity and mobility,
and the defects produced by grain boundaries and surfaces. These experiments were done in slowly deposited polycrystalline
CdS thin films, and at temperatures as 130$^{0}$C and 180$^{0}$C. The mobility is found to be strongly dependent of grain size,
and function of substrate temperature as well. They found a critical point that the grain boundary potential varies with
deposition rate. In a second paper\citep{LL72II} they investigated how the stacking-faults along with grain-boundary
affecting surface scattering impact greatly conductivity.

   The electrical conductivity and mobility of polycrystalline CdS thin films (prepared by spray pyrolysis) at dark and under illumination
were performed to correlate with thicknesses.\citep{FA09} Point defects play a vital role in this case.

    A photoluminescence technique reveals the inhomogeneous distribution of localized defect states of polycrystalline semiconductors
(for examples, CdS and CdTe).\citep{KA13} These defects are responsible for long-range photoconductivity through multiple grain boundaries.

  \citet{RB68} explained that polycrystalline CdS can be treated as a series connection between grain interior and boundary,
and illustrated to be inhomogeneous electronic conductivity. In Bube's model, the grain interior and boundary contain high and low conductivity
respectively. Even though carrier concentration and mobility are both competitive mechanisms, still it is not resolved which 
one becomes more dominant and under what condition.

 The electrical conductivity of thin-film semi-conductors is hard to
measure in the far-infrared or sub-millimeter regime.  Sources, both
pulsed (laser-based) or steady state (either laser photo-mixing or 
high harmonic generation of microwaves) are relatively weak, and the
film thicknesses of interest in photovoltaics are much less than the
wavelength of the probe beam.  Nevertheless, we want to understand the
conductivity behavior of CdS films in the sub-millimeter regime
since this is where electronic and optical properties strongly overlap.

   In this paper, we measure THz conductivity at room temperature. We use perturbation of an open hemispherical
cavity resonator\citep{RR13} to probe complex dielectric constants of CdS thin films, unannealed and annealed.
This method is also able to observe laser-induced photoexcitation measurements.\citep{RR14}
Even though in this sub-THz range the phonon-mediated mechanisms are predominant,\citep{zallen} we confirm the presence of THz 
electrical conductivity in CdS thin films.

 The CdS films were thermally evaporated 
in a vacuum chamber. CdS pieces were placed in a 6 mm quartz crucible 
that was heated by a tungsten filament. The vacuum was approximately 3 X 10$^{-6}$ torr. 
The substrate temperature was approximately 150$^{0}$C.

 We measure DC resistivity of CdS thin films with a van der Pau type device where
 a Lakeshore 7506 Hall system with an excitation current of 10 nA is implemented for such DC measurements.  

  \citet{RR13, RR14} describe the open cavity method for measuring complex dielectric properties of thin films in detail. In this experiment,
we repeated the whole process (of putting the sample into the cavity and taking out) for five times to estimate the measurement 
uncertainy. The pointwise standard deviation is calculated to be less than 1$\%$ of the measured spectra. The dark and room light measurements of 
dielectric constant ($\epsilon_{re}$) and conductivity ($\sigma_{re}$) for both
(unannealed and annealed) CdS films give the identical values. Moreover, between 100-180 GHz, the conductivity remains constant in frequency.
Due to absence of $\nu^{2}$-dependence, this does not represent any phonon vibrations or damping.
Therefore, at room temperature, this is an indication of thermally-activated electronic conductivity.

  The classical Drude model for frequency-dependent (real part of) conductivity is expressed by
\begin{equation}
 {\sigma}_{re}(\nu) =  \left\{ \frac{\sigma_{0}}{(1 + (2\pi\nu\tau^{2})}\right\}.\label{equ:Drude model}
\end{equation}     
where, $\tilde{\sigma}$ is the complex conductivity, $\sigma_{0}$ is
the dc conductivity and $\tau$ is the average collision time
of electrons.  It is a macroscopic parameter containing
both trap-capture time and also release-time.

The DC conductivity is defined by
\begin{equation}
 \sigma_{0} =  n_{e}^{(0)} e \mu_{e}^{(0)}.\label{equ:dcondvt}
\end{equation} 
With $n_{e}^{(0)}$ is the constant electron-concentration (holes are
ignored because of low-mobility and the films are typically n-type), $e$ is the electronic charge and
$\mu_{e}^{(0)}$ is the constant electron mobility. The relation between
the frequency-dependent electron mobility, $\mu_{e}(\nu)$ and
effective electron mass $m^{*}(\nu)$ is
$\mu_{e}(\nu)$=($e/m^{*}(\nu)\gamma(\nu)$), where,
$\gamma=1/\tau$, defined as carrier(electron) damping rate. Now,
the dispersion in $n_{e}(\nu)$, $n_{e}(\nu)= n_{e}^{(0)}f(\nu)$,
and in $\mu_{e}(\nu)$, $\mu_{e}(\nu)= n_{e}^{(0)}g(\nu)$, make them harder
to calculate. To do that, we have to find the frequency-dependence in
$f(\nu)$ and $g(\nu)$. 

\begin{figure}
  \includegraphics[width=65mm]{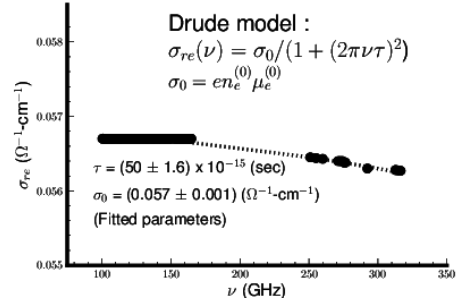}
  \caption{Sub-THz (100-312 GHz) data of conductivity of unannealed CdS.}
\label{fig:1}
\end{figure}

\begin{figure}
 \includegraphics[width=65mm]{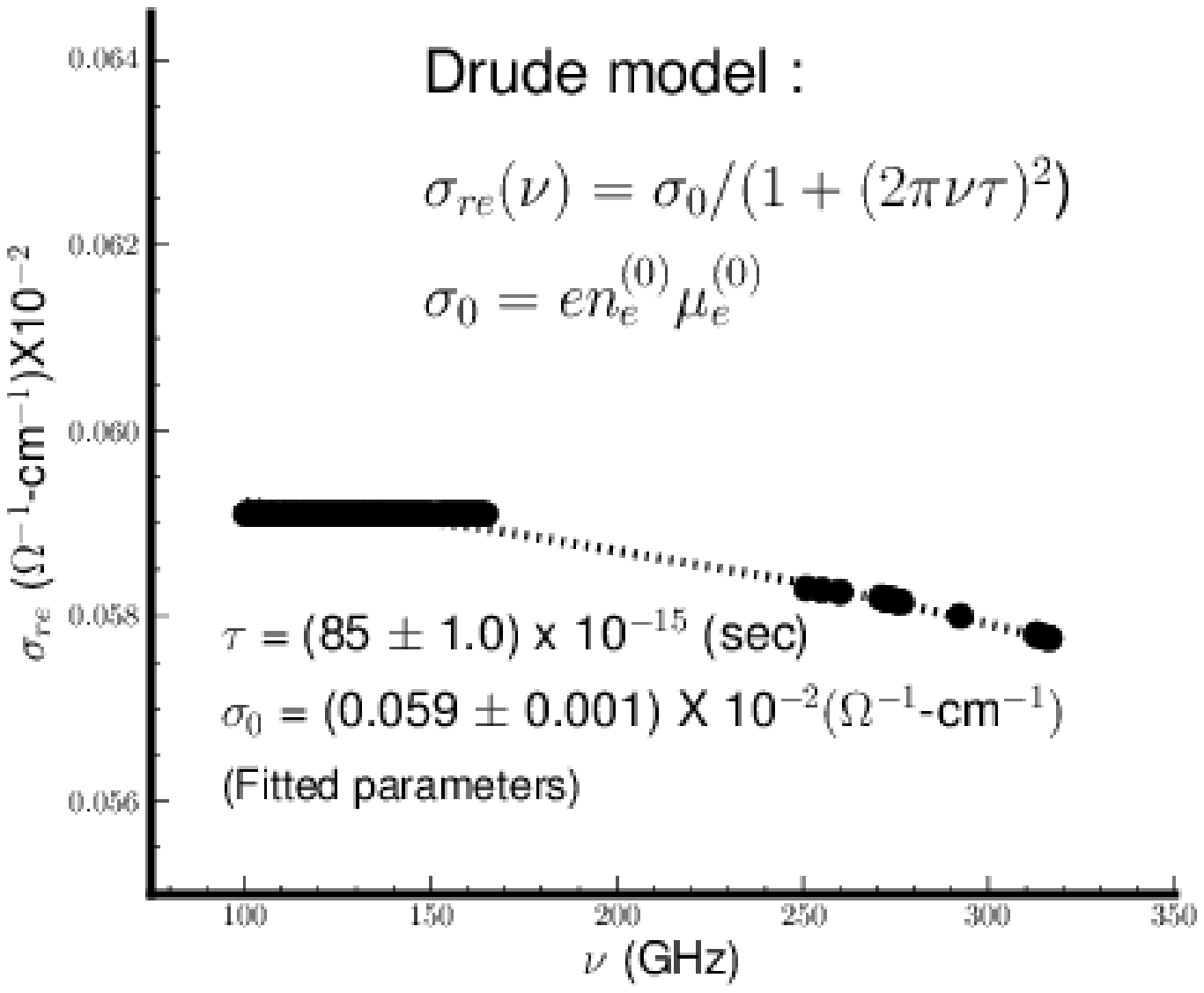}
   \caption{Sub-THz (100-312 GHz) data of conductivity of annealed CdS.}
\label{fig:2}
\end{figure}

\begin{figure}
 \includegraphics[width=65mm]{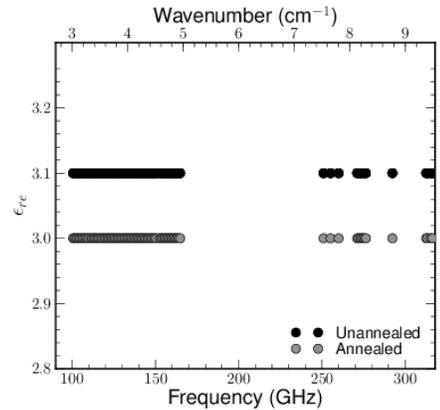}
   \caption{ Real part of dielectric constant of both unannealed and annealed CdS between 100 and 312 GHz.}
\label{fig:3}
\end{figure}

Figure \ref{fig:1}, \ref{fig:2} and \ref{fig:3}, shows that the experimental data and 
the model described by Eq.(\ref{equ:Drude model}). For unannealed CdS films, the fit parameters, $\tau$ = (50.0 $\pm$ 1.6) x 10$^{-15}$ sec., and
$\sigma_{re}^{(0)}$ = (0.057 $\pm$ 0.001) $\Omega^{-1}$ cm$^{-1}$; for annealed one, $\tau$ = (85.0 $\pm$ 1.0) x 10$^{-15}$ sec., and
$\sigma_{re}^{(0)}$ = (0.059 $\pm$ 0.001) X 10$^{-2}$ $\Omega^{-1}$ cm$^{-1}$ are
extracted from fitted-model.

\begin{table*}
\caption{\label{table1}Electronic conductivity profiles of unannealed and annealed CdS thin films (from the fits):\\
                       ($^{*}$ indicates femtosecond: 1 fs = 10$^{-15}$ sec)}
\begin{ruledtabular}
\begin{tabular}{cccc}
 CdS-samples  & Collision time  & DC conductivity           &Carrier concentration\\
              & $\tau$          & $\sigma_{0}$              & n$_{e}^{(0)}$       \\                                                                           
              & (fs)$^{*}$      &($\Omega^{-1}$ cm$^{-1}$)  & (cm$^{-3}$)          \\ \hline
 Unannealed   & (50.0 $\pm$ 1.6)            & (0.057 $\pm$ 0.001)                    & (3.65 $\pm$ 0.01) X 10$^{16}$ \\
 Annealed     & (85.0 $\pm$ 1.0)           & (0.00059 $\pm$ 0.00001)                  & (3.70 $\pm$ 0.01) X 10$^{14}$
\end{tabular} 
\end{ruledtabular}
\end{table*}

   The upper limit of electron mobility, $\mu_{e}^{(0)}$ $\simeq$ 10 cm$^{2}$V$^{-1}$sec$^{-1}$ at room temperature is reported.\citep{CMB70}
The formula is used for calculating $\mu_{e}^{(0)}$, is given as $\mu_{e}^{(0)} \simeq (ea^{2}/6k_{B}T) \nu_{e}$, based on 
the models.\citep{MC70,MT69} Here, $a$ is interatomic seperation and $\nu_{e}$ is the frequency of an electron, typically $\sim$ 
10$^{15}$ sec$^{-1}$. Using Eq.(\ref{table1}), the carrier concentrations, n$_{e}^{(0)}$, in unannealed and annealed samples are
(3.65 $\pm$ 0.01) x 10$^{16}$ cm$^{-3}$ and (3.70 $\pm$ 0.01) x 10$^{14}$ cm$^{-3}$, respectively, at room temperature.
Therefore, unannealed CdS contains $\sim$ 10$^{2}$ more carriers (per cm$^3$) than that of annealed one.

   Annealing makes the grain size larger and the depletion zone is wider including grain boundaries. This depletion layer contains
negligible carrier conentrations, and due to higher potential barrier, the conductivity of electrons
decrease at room temperature (without any external excitations such as lasers).\citet{ML12} mentioned that CdS thin films or nano belts are typically n-type due to sulfur vacancies on the surfaces. These are apparently active sites where the adsorptions of oxygen/water occur. Two depletion layers near both surfaces of nanobelts are created because of capturing intrinsic free electrons.
This process of trapping carriers can extend the width of the depletion layers during annealing of CdS thin films. The conductivity is extremely low due to lack of carriers which is also indicated by high resistivity (by DC conductivity measurements).
On the other hand, there is no clear demarcation between grain interior and boundary in an unannealed one.
Therefore, the space-charge region where the depletion region resides, is small and results in a higher
carrier concentration (than that of annealed sample). This also enhances electrical conductivity due to low
potential barrier.
  In higher frequency regions (250 - 320 GHz), the annealed sample shows slower decaying- electrical conductivity even though the average collision time is faster than that of the unannealed one; this is because it contains much lower carrier concentrations due to backscattering or localizations. 

    This work was supported by the US Department of Energy
(Basic Energy Science) under grant DE-FG02-09ER16018.  The authors would like to
thank Dr. Joe Beach, at Colorado School of Mines, for preparing these CdS thin films by thermally sublimation method.
We are very thankful to Dr. Tim R. Ohno for his DC conductivity  measurements on these samples and invaluable discussions
on the grain size and electrical conductivity of CdS/CdTe type materials.

\end{document}